\documentclass[useAMS,usenatbib]{mn2e}
\usepackage{hyperref}
\usepackage{amssymb,esint}
\usepackage{graphicx}
\usepackage{epstopdf}
\usepackage{times}
\newcommand{\bea}{\begin{eqnarray}}
\newcommand{\eea}{\end{eqnarray}}
\newcommand{\beq}{\begin{equation}}
\newcommand{\eeq}{\end{equation}}

\title[Gpc `structures' do not violate homogeneity]{Seeing patterns in noise: Gigaparsec-scale `structures' that do not violate homogeneity}
\author[S. Nadathur]{Seshadri Nadathur$^1$\\
$^1$Fakult\"at f\"ur Physik, Universit\"at Bielefeld, Postfach 100131, D-33501 Bielefeld, Germany}
\begin{document}

\date{\today}

\pagerange{\pageref{firstpage}--\pageref{lastpage}}

\label{firstpage}

\maketitle

\begin{abstract}
\citet{Clowes:2012pn} have recently reported the discovery of a Large Quasar Group (LQG), dubbed the Huge-LQG, at redshift $z\sim1.3$ in the Data Release 7 quasar catalogue of the Sloan Digital Sky Survey. On the basis of its characteristic size $\sim500$~Mpc and longest dimension $>1$ Gpc, it is claimed that this structure is incompatible with large-scale homogeneity and the cosmological principle. If true, this would represent a serious challenge to the standard cosmological model. However, the homogeneity scale is an average property which is not necessarily affected by the discovery of a single large structure. I clarify this point and provide the first fractal dimension analysis of the DR7 quasar catalogue to demonstrate that it is in fact homogeneous above scales of at most $130\,h^{-1}$~Mpc, which is much less than the upper limit for $\Lambda$CDM. In addition, I show that the algorithm used to identify the Huge-LQG regularly finds even larger clusters of points, extending over Gpc scales, in explicitly homogeneous simulations of a Poisson point process with the same density as the quasar catalogue. This provides a simple null test to be applied to any cluster thus found in a real catalogue, and suggests that the interpretation of LQGs as `structures' is misleading.
\end{abstract}

\label{firstpage}

\maketitle



\section{Introduction}
\label{section:intro}

A fundamental assumption of the standard $\Lambda$ Cold Dark Matter ($\Lambda$CDM) cosmological model, and indeed of all cosmological models based on a Friedmann-Robertson-Walker (FRW) metric, is that the Universe is close to homogeneous and isotropic. This means that properties of the Universe such as the matter density or the number density of galaxies should be invariant of spatial position. This is self-evidently not true on small scales and late times, where the distribution of matter is highly inhomogeneous and fluctuations are large. It is assumed that when viewed on larger scales, fluctuations should become smaller, and above a certain scale ($\sim100\,h^{-1}$~Mpc in the standard $\Lambda$CDM cosmology) they should be small enough to be negligible. 

Clearly such a statement is somewhat ambiguous, in that it depends on what size of fluctuation is regarded as negligible. Indeed the standard inflationary cosmology predicts fluctuations in the gravitational potential of similar amplitude on all scales, meaning that fluctuations in the matter density also do not go precisely to zero at any scale.
 
From a theoretical perspective, it may be interesting to ask whether the late-time inhomogeneities can affect the evolution of average quantities through the `backreaction mechanism' \citep[e.g.][]{Buchert:1999er,Ellis:2005uz,Li:2007ci}, rendering the exactly homogeneous and isotropic FRW models insufficient.\footnote{In such a scenario, a less stringent version of the cosmological principle, postulating statistical homogeneity and isotropy but allowing for large perturbations away from an FRW metric, can be adopted.} This is still an open area of research; see \citet{Rasanen:2011ki,Buchert:2011sx} for recent reviews. From an observational perspective, the question is instead one of consistency: do the observed density fluctuations at different scales agree with the expectations in the standard cosmological model?

For fluctuations in the dark matter density field, such a question can only be addressed indirectly, for instance through measurement of the effect of large dark matter inhomogeneities on the cosmic microwave background via the integrated Sachs-Wolfe (ISW) effect of isolated structures. Indeed there is evidence of tension between the observed and expected ISW signals of the rarest structures at scales of $\gtrsim100h^{-1}$~Mpc \citep*[see for instance][]{Granett:2008ju, Hunt:2008wp, Nadathur:2011iu, Flender:2012wu,HernandezMonteagudo:2012ms}, which may indicate that dark matter inhomogeneities on such scales are larger than expected.

Inhomogeneities in the distribution of visible matter can be studied more directly. Given any large redshift catalogue of visible objects that trace the matter density field, two distinct approaches may be taken to the question of testing whether it is compatible with $\Lambda$CDM (or any other FRW cosmological model). 

The first approach is to determine whether the catalogue as a whole is homogeneous on large scales, and if it is, whether the onset of homogeneity thus measured occurs at the scales expected in $\Lambda$CDM. This is usually done using a fractal analysis based on the `counts-in-spheres' measurement of the average number of objects $N(<R)$ contained within spheres of radius $R$ centred on an object in the catalogue. This average scales as $N(<R)\propto R^{D_2}$, which serves to define the correlation dimension $D_2(R)$. For a homogeneous distribution, $N(<R)$ should scale as $R^3$, i.e. $D_2=3$. The scale above which a given catalogue satisfies this property to within the desired precision may be referred to as the homogeneity scale. \citet*{Yadav:2010cc} provide a conservative upper limit of $R_\mathrm{H}<260\,h^{-1}$~Mpc for the scale by which this transition should be observed in the $\Lambda$CDM model; in practice the scale is expected to be much smaller.

Historically, there was some debate over whether such a transition to homogeneity had been observed in shallow redshift surveys that were not ideally suited to this test \citep[see][and references within for a summary]{Scrimgeour:2012wt}. Using the SDSS Luminous Red Galaxy (LRG) sample \citep{Eisenstein:2001cq}, which is better suited to such tests, \citet{Hogg:2004vw} found a homogeneity scale of $R_\mathrm{H}\sim70\,h^{-1}$~Mpc. Subsequently \citet{Scrimgeour:2012wt} showed (using a slightly different definition of $R_\mathrm{H}$) that subsamples of the WiggleZ survey \citep{Drinkwater:2009sd} are compatible with homogeneity at scales above $70\lesssim R_\mathrm{H}\lesssim90\,h^{-1}$~Mpc. On the other hand, some authors claim to find no large-scale homogeneity in other catalogues \citep*{Labini:2009nx,Labini:2009mv, Labini:2011dv}. Actually, this is a basic test of homogeneity which should be applied to every redshift catalogue independently. This is because even if the matter distribution of the Universe is homogeneous, the distribution of galaxies in an inappropriately chosen sample may not be. Large-scale homogeneity of a given catalogue is however a necessary precondition for other statistical quantities determined from it, such as the two-point correlation function, to be meaningful \citep{Gabrielli:2005bk}.

The second approach to testing compatibility with $\Lambda$CDM, which may usefully be applied even to a catalogue passing the first test, is to search for specific rare structures or density fluctuations within it. The properties of such structures, if found, can then be carefully compared with the predictions for their existence in $\Lambda$CDM. This approach is independent of the fractal analysis, in the sense both that it is possible to have individual structures consistent with a $\Lambda$CDM cosmology that extend over scales larger than the homogeneity scale, and that structures which contradict the detailed predictions of $\Lambda$CDM need not affect the overall homogeneity of the catalogue. This is because $N(<R)$ and $D_2(R)$ are \emph{average} quantities, so the homogeneity scale is a property of the catalogue considered as a whole and -- for a large enough catalogue -- only weakly affected by individual fluctuations.

Some examples of luminous superclusters found in the 2dF Galaxy Redshift Survey and the SDSS Data Release 4 have been claimed to be in some tension with predictions \citep{Einasto:2006es,Einasto:2006si,Einasto:2006mm}. Studies of other structures in the 2dFGRS \citep*{Yaryura:2010uf,Murphy:2010bb} also hint towards tension with theoretical expectations, although it is not clear whether the discrepancy is due to failings of the $\Lambda$CDM cosmological model, or to models of galaxy formation.

In following the second approach and testing the standard cosmology through observations of individual structures, however, care must be taken in the correct quantification of the likelihood of their existence in the standard model, which will in general depend on the definition of what constitutes a `structure'. For instance, the Sloan Great Wall \citep[SGW;][]{Gott:2003pf} -- a filamentary structure identified in the SDSS galaxy distribution that extends over more than 400 Mpc -- has been suggested to be extremely unlikely in $\Lambda$CDM \citep{Sheth:2011gi}, yet \citet{Park:2012dn} find that structures as large or larger are in fact not unusual in large $N$-body cosmological simulations.

Recently, however, \citet{Clowes:2012pn} have reported the discovery of an even larger structure in the SDSS Data Release 7 quasar catalogue \citep[DR7QSO;][]{Schneider:2010hm}, identified through the use of a three-dimensional single-linkage hierarchical clustering algorithm. Known as the Huge-LQG, this structure is reported to have a characteristic size (defined as volume$^{1/3}$) of $\sim500$~Mpc, and a longest dimension in excess of $1$ Gpc, making it far larger than the SGW. It is claimed that the existence of such a structure is incompatible with the \citet{Yadav:2010cc} upper limit to the scale of homogeneity, and thus challenges the cosmological principle.\footnote{Actually the cosmological principle, understood in the sense of requiring only statistical homogeneity and isotropy as discussed above, makes no statement about the \emph{scale} above which this homogeneity should be achieved. The implied challenge of the Huge-LQG is really specifically to the $\Lambda$CDM model.} If true, this would be a very significant discovery.

However, although the quoted dimensions of the Huge-LQG are at first sight surprisingly large, it is not at all clear what implications it has for the question of the scale of homogeneity of the catalogue as a whole. It is also not clear how unlikely the Huge-LQG actually is in $\Lambda$CDM, nor what role the clustering algorithm used in its detection has in assessing this likelihood.

These are the questions addressed in this paper. To do so, I first apply a fractal analysis to the DR7QSO catalogue and demonstrate that it is in fact entirely compatible with homogeneity at large scales. As already mentioned, the exact definition of `the scale of homogeneity' is somewhat ambiguous, and in any case the rather sparse nature of the quasar catalogue (mean nearest-neighbour distances are $\sim75$~Mpc) means it is not well-suited to a precise determination; however, $R_\mathrm{H}$ is certainly less than $\sim130\,h^{-1}$~Mpc. On the other hand, the extremely large volume of the DR7QSO catalogue and its relatively simple geometry mean that the fractal analysis can be applied without requiring additional prior assumptions about the large-scale homogeneity that is the subject of the test. This was not the case for the analysis by \citet{Scrimgeour:2012wt}, due to the use of a correction to number counts for incompletely sampled spheres that presupposed homogeneity, though the effect of this was argued to be small. It was also not the case for \citet{Hogg:2004vw}, where, although no completeness corrections were used, $N(<R)$ counts were normalized relative to those expected in a homogeneous distribution.

I then investigate the role of the hierarchical clustering algorithm used by \citet{Clowes:2012pn} to identify the Huge-LQG by applying it to $10,000$ homogeneous simulations of a Poisson point process with the same number density of points as the DR7QSO catalogue, and finding the largest `cluster' in each. I  examine the dependence of the cluster size on the minimum single-linkage length cutoff used to define a cluster and provide a simple fit in terms of extreme value statistical distributions. Clusters of points as large as the Huge-LQG or larger -- both in membership and in spatial extent -- are found in about $8.5\%$ of these simulations. This shows that the statistical significance attributed to the discovery of the Huge-LQG is vastly overstated, and that it is entirely compatible with random expectations. This conclusion applies even more strongly to other smaller quasar groups reported in the past \citep{Clowes:1991,Clowes:2011eb}. In light of this, I suggest that it is misleading to refer to these quasar groups as `structures' at all.

\begin{figure*}
\begin{center}
\includegraphics[width=175mm]{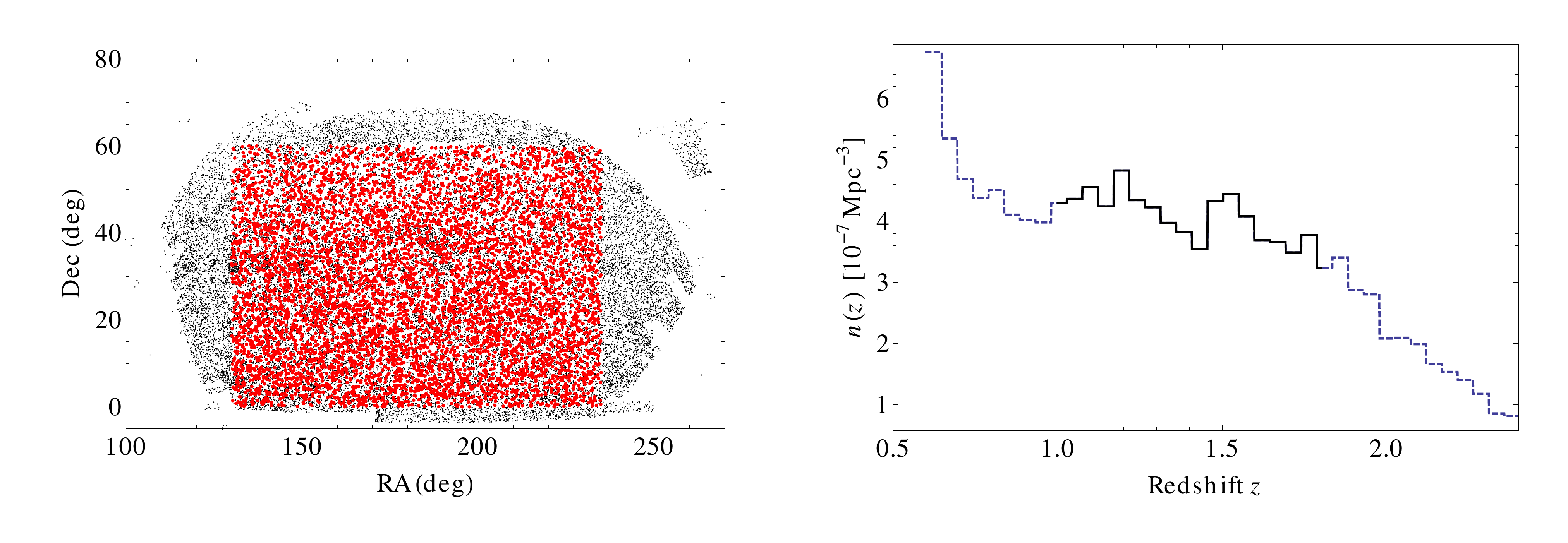}
\caption{\emph{Left panel}: The small black points show the right ascension and declination coordinates of DR7QSO quasars around the North Galactic Pole. The larger, red points show the coordinates of those quasars which are part of the SCR subsample. For display purposes, only one third of the quasars in each group, selected at random, are shown in this figure.
\emph{Right panel}: The redshift distribution of DR7QSO quasars that satisfy the $(\alpha,\delta)$ cuts applied in the text. The blue dashed line shows the comoving number density for all quasars and the solid black line for those in the redshift range $1.0\leq z\leq1.8$ used to define the SCR.} 
\label{figure:RADec}
\end{center}
\end{figure*}

In Section~\ref{section:DR7QSO} I briefly describe the criteria used to select a suitable subsample from the DR7QSO catalogue and some of its properties. Section~\ref{subsec:counts} describes the fractal analysis test for large-scale homogeneity, and different definitions of the average `scale of homogeneity'; Section~\ref{subsec:clustering} discusses some aspects of the hierachical clustering approach to finding structures. I describe the methodology used in this paper in Section~\ref{section:methodology} and the results in Section~\ref{section:results}. The implications for homogeneity and the interpretation of LQGs as `structures' are discussed in Section~\ref{section:conclusion}.

For calculation of cosmological distances, I assume a flat Universe with the parameter values $\Omega_M=0.27$, $\Omega_\Lambda=0.73$ and $H_0=70$ kms$^{-1}$Mpc$^{-1}$. All distances quoted are comoving distances.

\section{The SDSS quasar catalogue}
\label{section:DR7QSO}

In this work I use the SDSS DR7QSO catalogue of $105,783$ quasars \citep{Schneider:2010hm}. The majority of these quasars were identified as part of the SDSS Legacy Survey, which consists of a large contiguous area around the North Galactic Pole (known as the North Galactic Cap or NGC), and some narrow stripes near the celestial equator. Some additional quasars found on a series of `special plates' complete the rest of the catalogue. In total the catalogue covers a region of $\simeq9380$ deg$^2$ on the sky.

The DR7QSO catalogue does not constitute a statistical sample due to changes in the target strategy at different redshifts.  However, if focusing on only the low-redshift ($z\lesssim2$) quasars, a satisfactorily homogeneous selection can be achieved by limiting the $i$-band magnitude to $i\leq19.1$ \citep{Schneider:2010hm, Richards:2006tc,VandenBerk:2005di}. This is also the selection criterion applied by  \citet{Clowes:2011eb} and \citet{Clowes:2012pn}, and is therefore adopted here. Following these papers, this analysis also considers only those quasars in the redshift range $1.0\leq z\leq1.8$. In order to avoid the complications of jagged boundaries for calculating the counts-in-spheres test and comparison with simulated homogeneous distributions, the sample is further restricted to the contiguous region within the NGC bounded by right ascension $130^\circ\leq\alpha\leq235^\circ$ and declination $0^\circ\leq\delta\leq60^\circ$. 

I shall refer to the subsample thus defined as the Simple Contiguous Region (SCR). It contains $18,722$ quasars, and completely encompasses the Huge-LQG of \citet{Clowes:2012pn}, the smaller U1.28 and U1.11 quasar groups of \citet{Clowes:2011eb}, as well as the `control region' designated A3725 by those authors. Figure~\ref{figure:RADec} shows the angular distribution of these quasars in $(\alpha,\delta)$ coordinates superimposed on the distribution of all quasars around the North Galactic Pole, and their comoving number density as a function of $z$. The redshift distribution over the range $1.0\leq z\leq1.8$, though not completely flat, is sufficiently uniform for our purposes.

\begin{figure}
\begin{center}
\includegraphics[width=84mm]{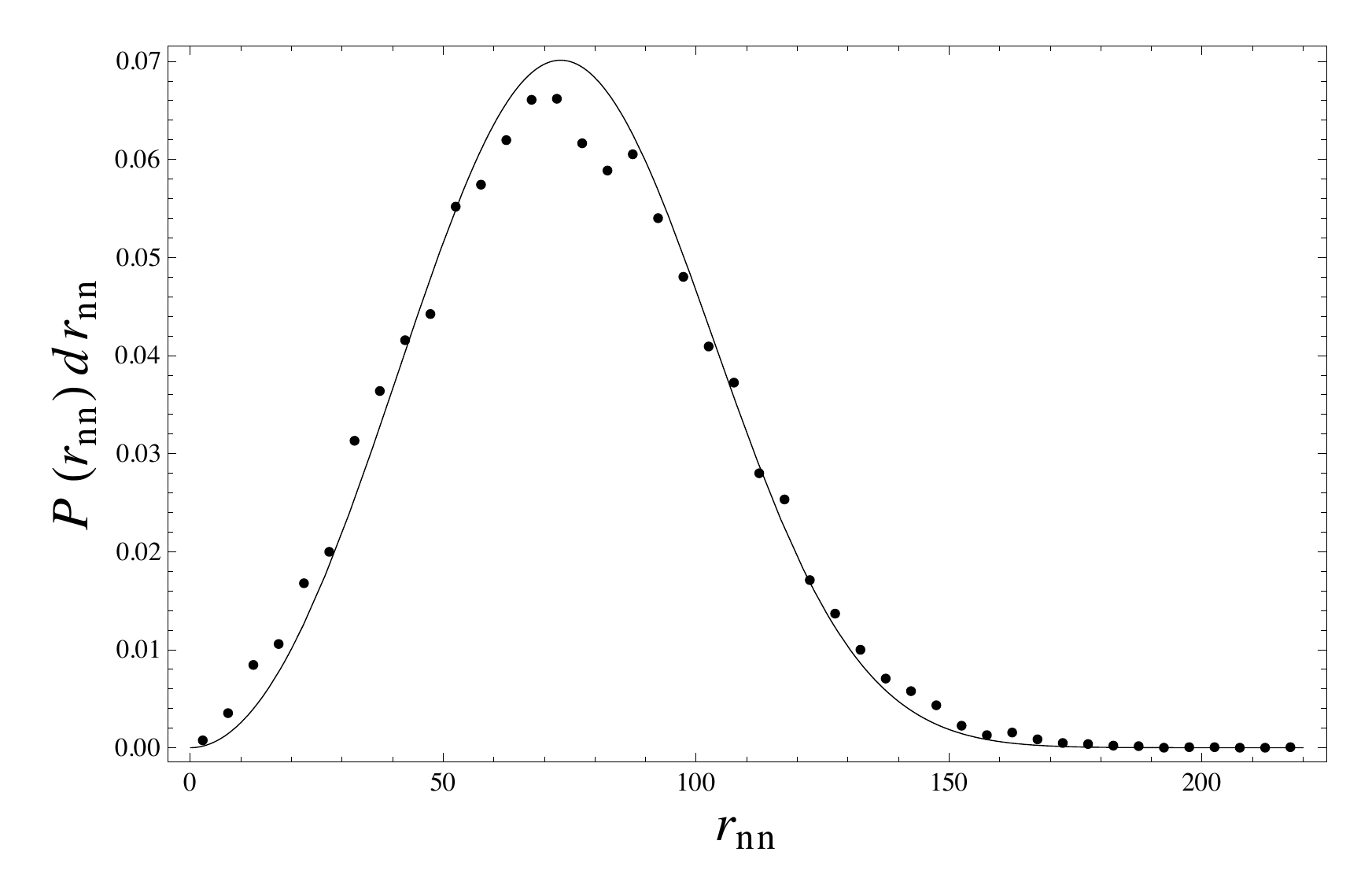}
\caption{The distribution of nearest-neighbour distances for quasars in the SCR subsample. The black points show the relative numbers in bins of $5$~Mpc width. The solid line shows the theoretical expectation for a homogeneous Poisson distribution of the same number points in the same volume.} 
\label{figure:NNdist}
\end{center}
\end{figure}

Because of its high central redshift, depth and wide angular extent, the SCR occupies a very large comoving volume, $\sim46$~Gpc$^3$. This makes it well-suited to testing the homogeneity of the quasar distribution on extremely large scales. It is however extremely sparse, with a mean nearest-neighbour separation of $\bar{r}_\mathrm{nn}=74.5$~Mpc. This is remarkably close to the mean nearest-neighbour distance for a homogeneous Poisson distribution of points with the same mean density, $\bar{r}^\mathrm{P}_\mathrm{nn}\equiv 0.55\left(N/V\right)^{-1/3}=74.3$~Mpc. Figure~\ref{figure:NNdist} shows the distribution of nearest-neighbour distances for the SCR, and the expectation for the Poisson case; despite expected broadening of the tails due to clustering effects, the two are indeed very similar. 

Although the SCR encompasses the Huge-LQG, in performing their algorithmic search for quasar clusters, \citet{Clowes:2012pn} did not impose the tighter angular cuts applied here but instead included all quasars in the NGC region that satisfied the redshift selection criterion. The mean nearest-neighbour distance for this larger set of NGC quasars is not much larger, $\bar{r}_\mathrm{nn}=75.2$~Mpc.

\section{Testing homogeneity with redshift catalogues}
\label{section:homogeneity}

\subsection{Fractal analysis}
\label{subsec:counts}

The simplest test of homogeneity that can be applied to any point set is based on the average of the number of neighbouring points $N_i(<R)$ contained within a sphere of radius $R$ centred on the $i$th member of the point set, with the requirement that the entire sphere lies within the distribution of points:
\beq
\label{eq:N<R}
N(<R) = \frac{1}{M}\sum_{i=1}^{M}N_i(<R)\;,
\eeq
where $M$ is the number of sphere centres. For a homogeneous distribution $N(<R)\propto R^{D}$, where $D$ is the number of dimensions, three in this case. The correlation dimension $D_2(R)$ is calculated as the derivative 
\beq
\label{eq:D2}
D_2(R)=\frac{d\;\mathrm{ln}\,N(<R)}{d\;\mathrm{ln}\,R}\;,
\eeq
and quantifies the deviation from this homogeneous scaling.

For any given catalogue of objects that trace the matter density of the Universe, $N(<R)$ can be related to the two-point correlation function $\xi(r)$ by
\beq
\label{eq:Nandxi}
N(<R) = \bar{\rho}\int_0^R \left(1+b^2\xi(r)\right)4\pi r^2 dr\;,
\eeq
where $\bar\rho$ is the mean matter density and $b$ is the bias of the tracer population. Note that the relationship in eq.~(\ref{eq:Nandxi}) requires the assumption that the homogeneous background exists at large scales, as it is only under this assumption that $\bar\rho$ and $\xi(r)$ are meaningful quantities. $N(<R)$ can however be calculated for any catalogue without assuming homogeneity.

As can be seen from eqs.~(\ref{eq:D2}) and~(\ref{eq:Nandxi}), clustering effects mean that even in the standard $\Lambda$CDM model, $D_2<3$ on small scales. Indeed it is known that on small scales the two-point correlation function measured in galaxy surveys is well approximated by a power-law form
\[
\label{eq:powerlawxi}
\xi(r) = \left(\frac{r_0}{r}\right)^\gamma\;,
\]
where $r_0\simeq5h^{-1}$~Mpc and $\gamma\sim1.8$ \citep[e.g.][]{Peebles:1993bk}. On larger scales, if the galaxy sample in question approaches homogeneity, $D_2$ should asymptotically approach $3$. However, the precise definition of the scale above which homogeneity is achieved is a subjective question, which depends on the criterion by which differences from homogeneous scaling are judged.

\citet{Gabrielli:2005bk} use individual $N_i(<R)$ rather than the average $N(<R)$, and define the homogeneity scale as the value of $\lambda_0$ such that
\beq
\label{eq:lambda0}
\left| \frac{3N_i(<R)}{4\pi R^3}-\rho_g \right|<\rho_g\;\;\forall R>\lambda_0,\;\forall i\in\{1,2,\ldots M\},
\eeq
where $\rho_g=N/V$ is the overall density of points in the set. This definition is extremely restrictive, since the condition must be satisfied for \emph{all} centres. It therefore also has the disadvantage that the existence of rare fluctuations means that $\lambda_0$ must increase as the number of centres $M$ grows, so that the homogeneity scale of a galaxy catalogue increases with its size. 

\citet*{Bagla:2007tv} suggest instead defining the scale of homogeneity as being the scale at which the average correlation dimension is consistent with the homogeneity value within one standard deviation, i.e. $\vert D_2(R)-3\vert<\sigma_{\Delta D_2}$. However, such a definition is also survey-dependent, since the error bars on the data depend on the survey size, details of its geometry and selection function, as well as shot noise and cosmic variance effects. Considering only the latter two contributions to $\sigma_{\Delta D_2}$, \citet{Yadav:2010cc} find an upper limit to the homogeneity scale of $260h^{-1}$~Mpc. Using this definition the scale actually measured in a real survey will necessarily be smaller. Indeed, \citet{Hogg:2004vw} appear to use a similar criterion applied to $N(<R)$ determined for the SDSS LRG sample \citep{Eisenstein:2001cq}, and find scaling compatible with homogeneity at scales $\gtrsim70\,h^{-1}$Mpc.

\citet{Scrimgeour:2012wt} choose instead to define the homogeneity scale as that scale above which a polynomial fit to either $N(<R)$ or $D_2(R)$ determined from the data crosses an arbitrary threshold, in this case taken to be $1\%$ away from the homogeneous value. Such a definition avoids the problem of survey-dependent errors, but depends instead on the bias of the tracer population and the survey epoch; for different subsamples of the WiggleZ survey they find values in the range $70\lesssim R_\mathrm{H}\lesssim90\,h^{-1}$~Mpc.

Both the latter two definitions of the homogeneity scale depend on the \emph{average} quantities $D_2(R)$ and $N(<R)$ determined over all sphere centres. For a large enough survey, this means that fluctuations about any small subset of sphere centres have little effect on the result. Therefore the existence of individual void or cluster structures in a galaxy or quasar catalogue cannot be used to make inferences about its large-scale homogeneity. Such individual structures \emph{would} affect the scale $\lambda_0$ defined in eq.~(\ref{eq:lambda0}); however, this definition is not commonly used in homogeneity studies.

\subsection{Hierarchical clustering}
\label{subsec:clustering}

A simple method of identifying structures in a point set such as the DR7QSO catalogue is to use a three-dimensional single-linkage hierarchical clustering algorithm, also sometimes called a percolation algorithm or a `friends-of-friends' (FOF) algorithm. In this method, points are grouped together by placing spheres of radius $L$ centred on each point of the catalogue. Overlapping spheres then constitute a `cluster', the membership or `richness' of each cluster being denoted by $k$.

This method has been used to search for clusters in several different astrophysical contexts, including \citet{Huchra:1982,Press:1982,Clowes:1991, Einasto:1997zd,Sheth:2011gi, Clowes:2011eb,Park:2012dn,Clowes:2012pn}. The advantage of such an algorithm is that it is independent of assumptions about the shape or morphology of the clusters. However, the interpretation of the results depends on appropriate choice of the linkage length $L$. 

One option is to choose $L$ to maximise the the fraction of clusters found that match some physical characteristics expected to correspond to those of real structures. Another is to maximise the number of clusters of $k>1$. By simply increasing $L$, one can certainly increase the likelihood of finding a large cluster of points, but this may not correspond to any physical structure. The probability of such false positive detections must be considered when specifically searching for large clusters.

To quantify this, we can parametrize $L$ in terms of the mean nearest-neighbour separation of points in the set $\bar{r}_\mathrm{nn}$:
\[
L=\beta \bar{r}_\mathrm{nn}.
\] 
For a homogeneous Poisson distribution of points, a critical percolation threshold exists above which infinite clusters (in practical terms, clusters which extend from one boundary of the volume in question to another) start to appear. This occurs at $\beta_c\simeq1.57$ \citep{Gayda:1974,Fremlin:1976}. 

The linkage length chosen by \citet{Clowes:2011eb} and \citet{Clowes:2012pn} is $L=100$~Mpc.  Given the values of $\bar{r}_\mathrm{nn}$ found in Section~\ref{section:DR7QSO}, this gives a value of $\beta$ that is at least $1.33$. Although this is below the critical threshold, the value appears quite large and clearly increases the probability of finding spurious large clusters in noise. Note here that the Huge-LQG consists of only $73$ quasars out of a total of $\sim19,000$ in the SCR subsample, so it does not have a particularly large membership. Therefore a careful estimation of the probability that such a cluster could be found in random noise is required.

\citet{Clowes:2012pn} attempt to do this by calculating the volume of the convex hull of spheres of radius $33$~Mpc (half the mean linkage length of member quasars of the Huge-LQG) placed at the $73$ member locations. This volume is called the convex hull of member spheres (CHMS) volume of the Huge-LQG, and is then compared with the average CHMS volume of $73$ uniformly distributed points placed in a box of volume such that the number density of points approximately matches that of the DR7QSO quasars, over $1000$ realisations. Based on this, the authors claim that the Huge-LQG represents a $3.81\sigma$ departure from random expectations.

However, such a comparison is essentially meaningless. It is hardly surprising that the $73$ members of the Huge-LQG occupy a smaller volume than the same number of uniformly distributed points, since the cluster-finding algorithm explicitly ensures that they constitute the most tightly linked group of $73$ quasars that could be selected from the full SCR subsample of $18,722$! Instead a sensible estimation of the probability that the Huge-LQG could arise from noise can only be made by comparing it to the largest cluster found by applying the same algorithm to a random catalogue of the \emph{same size and density}. This is done as described in the next section.

\begin{figure*}
\begin{center}
\includegraphics[width=176mm]{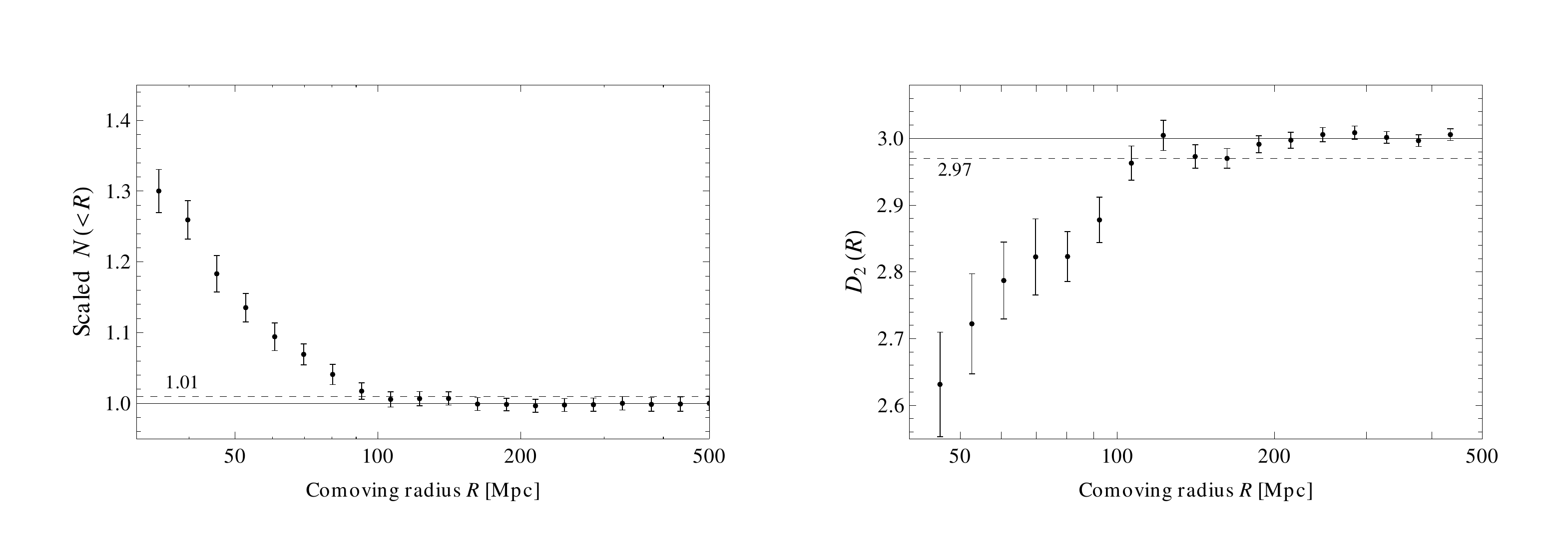}
\caption{\emph{Left panel}: Values of $N(<R)$ determined by applying the counts-in-spheres test to the SCR subsample for different sphere radii $R$. The values are scaled relative to the value at $R=500$~Mpc as described in the text. The solid horizontal line at the value 1 describes the expectation for homogeneous scaling; the dashed line shows a $1\%$ deviation from it.
\emph{Right panel}: The correlation dimension $D_2(R)$ determined from $N(<R)$ measurements. The solid horizontal line describes the expectation for homogeneous scaling and the dashed line shows a $1\%$ deviation from it.} 
\label{figure:NandD2}
\end{center}
\end{figure*}

\section{Methodology}
\label{section:methodology}

This section describes the methodology used in testing the SCR quasar subsample according to the two approaches described above. In order to apply these tests the redshift and angular coordinates of each quasar are first converted into comoving Cartesian coordinates 
\[
x=\chi \cos\delta \cos\alpha,\;y=\chi \cos\delta \sin\alpha,\;z=-\chi \sin\delta,
\]
where $\chi(z)$ is the comoving distance to redshift $z$, and $(\alpha,\delta)$ are the right ascension and declination coordinates of the quasar. Comoving distances are calculated for a $\Lambda$CDM model with parameter values stated above. This introduces an implicit and unavoidable prior assumption of homogeneity and isotropy. However, this is the only such assumption made in the analysis. If the quasar distribution truly were inhomogeneous in some way, one might reasonably expect that this would still be measurable using the fractal analysis (see \citealt{Scrimgeour:2012wt} for further discussion of this point).

\subsection{Determining $N(<R)$ and $D_2(R)$}

The first step in the fractal analysis of the SCR subsample is the determination of the average counts-in-spheres $N(<R)$ defined by eq.~(\ref{eq:N<R}). This is done at 21 logarithmically spaced values of R between 30 and 500 Mpc. At each radius, only those quasars are chosen as sphere centres for which the entire sphere is located within the boundaries used to define the SCR subsample. Other methods for correcting for boundary effects without restricting the number of sphere centres could also be used \citep{Martinez:1998,Pan:2001vw, Scrimgeour:2012wt} but these make further undesirable \emph{a priori} assumptions about the homogeneity or isotropy of the sample. Due to the size of the SCR volume, the restriction used here allows the use of a relatively large number of quasars as sphere centres even at large $R$ and so is adequate for our purposes.

Having obtained the $N(<R)$ values for different sphere radii $R$, $D_2(R)$ can be calculated from eq.~(\ref{eq:D2}) using a finite-difference approximation for the derivative. 

For convenience of visualisation, the $N(<R)$ values are scaled relative to the value at $R=500$~Mpc by dividing by a factor of $(R/500)^3N(<500)$.  This rescaling ensures that at $R=500$~Mpc, the scaled $N(<R)$ values must necessarily be equal to 1. However, if homogeneity is attained before this scale (as expected) then the scaled $N(<R)$ should approach 1 and stay at 1 above some smaller scale. Alternatively, the approach to homogeneity can be judged by the values of $D_2(R)$, which should approach 3 and stay at 3 for a homogeneous distribution. Note that this scaling procedure is different to those used in previous analyses \citep{Hogg:2004vw,Scrimgeour:2012wt} which introduced a further assumption of homogeneity.\footnote{Rescaling $N(<R)$ by the values expected in a homogeneous distribution, as done in these two papers, presupposes the existence of a mean density on the scale of the survey, ensuring that the rescaled value approaches unity on these scales. This would not affect the behaviour of $D_2(R)$, which could still be used to judge the approach to homogeneity; unfortunately \citet{Hogg:2004vw} did not consider this quantity.}

To estimate the errors in these measured values I use 100 realisations of a homogeneous Poisson distribution of the same number of points within the SCR volume, and determine $N(<R)$ and $D_2(R)$ for each realisation as before. The covariance matrix between radial bins $i$ and $j$ is calculated by
\beq
\label{eq:covmat}
C_{ij} = \frac{1}{n-1}\sum_{l=1}^n \vert(X_l(R_i)-\overline{X(R_i)}\vert \vert(X_l(R_j)-\overline{X(R_j)}\vert\,,
\eeq
where $X(R)$ is either $N(<R)$ or $D_2(R)$, the sum is over $n=100$ realisations and the bar denotes the average quantity determined over the realisations. The diagonal elements of this covariance matrix give the variance $\sigma^2$ at each radius.

Strictly speaking, the errors calculated using eq.~(\ref{eq:covmat}) are the errors that would be expected in a homogeneous distribution, and not those expected in the quasar catalogue below the homogeneity scale since gravitational clustering effects have been neglected. However, Figure~\ref{figure:NNdist} provides reason to believe that the effects of clustering in the SCR subsample are small; in addition, for the primary purpose of determining the scale above which the distribution is indistinguishable from a homogeneous one, considering the error bars for the homogeneous distribution is sufficient.

\subsection{Finding clusters in simulations}

In order to identify clusters of quasars according to the clustering algorithm used by \citet{Clowes:2012pn}, I make use of the 
Hierarchical Clustering package in {\small MATHEMATICA} to create a cluster hierarchy based on the distance matrix for the quasar coordinates. For an input maximum linkage length $L$, this hierarchy is then explored using a custom code to select the largest cluster by membership which satisfies the linkage length cutoff. If two clusters have the same number of members, the one with the smaller maximum linkage length is selected.

To check the functioning of the algorithm, I applied it to the SCR subsample with $L=100$~Mpc as used by \citeauthor{Clowes:2012pn}, and confirmed that it identified the 73 quasars of the Huge-LQG as reported. The maximum distance between any two quasars in this group is $D_\mathrm{max}\approx1076$~Mpc.

It is worth noting that when applied only to the quasars within the `control region' A3725, which is well separated from all previously reported LQGs, the same algorithm found \emph{another} giant quasar group, composed of 54 quasars and with $D_\mathrm{max}\approx730$~Mpc. On setting the linkage length $L=\bar{r}_\mathrm{nn}\sim74$ Mpc (i.e., $\beta=1$), the largest quasar group in the SCR sample consisted of 15 quasars situated around $(\alpha,\delta)\sim(207^\circ,27^\circ)$, with $D_\mathrm{max}\approx255$~Mpc. These results suggest that the Huge-LQG is not a particularly exceptional cluster.

Having thus checked the algorithm, I then applied it to $10,000$ realisations of a homogeneous Poisson point process, occupying the same region of space as the SCR (i.e. the same volume and geometry), and with the same mean number density of points. For the largest cluster found in each realisation, I recorded the number of members $k_\mathrm{max}$, the maximum point separation $D_\mathrm{max}$, and the volume of the convex hull (not CHMS) formed by the member points. This was done both with $\beta=1$, (i.e., the linkage length cutoff set to the mean nearest-neighbour distance), and with $\beta=1.33$. The latter value is the most generous estimate of the value used by \citeauthor{Clowes:2012pn}. In fact, given the slightly smaller $\bar{r}_\mathrm{nn}$ value for the SCR subsample compared to all NGC quasars, it corresponds to a value of $L$ slightly less than the $100$~Mpc used by those authors.

\section{Results}
\label{section:results}

\subsection{$N(<R)$ and $D_2(R)$}

Figure~\ref{figure:NandD2} shows the  behaviour of $N(<R)$ and $D_2(R)$ at different scales. It can clearly be seen that both show a clear approach to homogeneity at scales far below the maximum values probed. In particular, the scaled $N(<R)$ values are equal to 1 to within one standard deviation at all scales above $\sim100$~Mpc. The correlation dimension shows some additional small fluctuations at the $1\%$ level but is also consistent with the homogeneous value $D_2=3$ to within the error bars at scales $R\gtrsim180$~Mpc, (which corresponds to $R\gtrsim130\,h^{-1}$Mpc given the choice of $h$). Both $N(<R)$ and $D_2(R)$ are within $1\%$ of their homogeneous values (the criterion used by \citealt{Scrimgeour:2012wt} to determine homogeneity) at scales above $\sim100$~Mpc.

We can therefore conclude that the SCR subsample of the DR7QSO catalogue is compatible with homogeneity above at most $R\gtrsim180$~Mpc. The sparseness of the quasar distribution means it is not ideally suited to a precise determination of the scale of homogeneity (and such a scale is in any case not unambiguously defined), but it is certainly perfectly compatible with the \citeauthor{Yadav:2010cc} upper limit of $R_\mathrm{H}\sim370$~Mpc for a $\Lambda$CDM universe.

\subsection{Clusters from simulations}

A total of 849 of the $10,000$ Poisson simulations analysed with $\beta=1.33$ had a largest cluster with $k_\mathrm{max}\geq73$, meaning that  measuring by cluster membership the Huge-LQG is statistically distinct from clusters found in random noise at less than $92\%$ C.L. Expressed in the same terms as do \citet{Clowes:2012pn}, the significance of the departure from random expectations for the Huge-LQG is less than the $2\sigma$ level, a very different conclusion to theirs.

\begin{figure}
\begin{center}
\includegraphics[width=84mm]{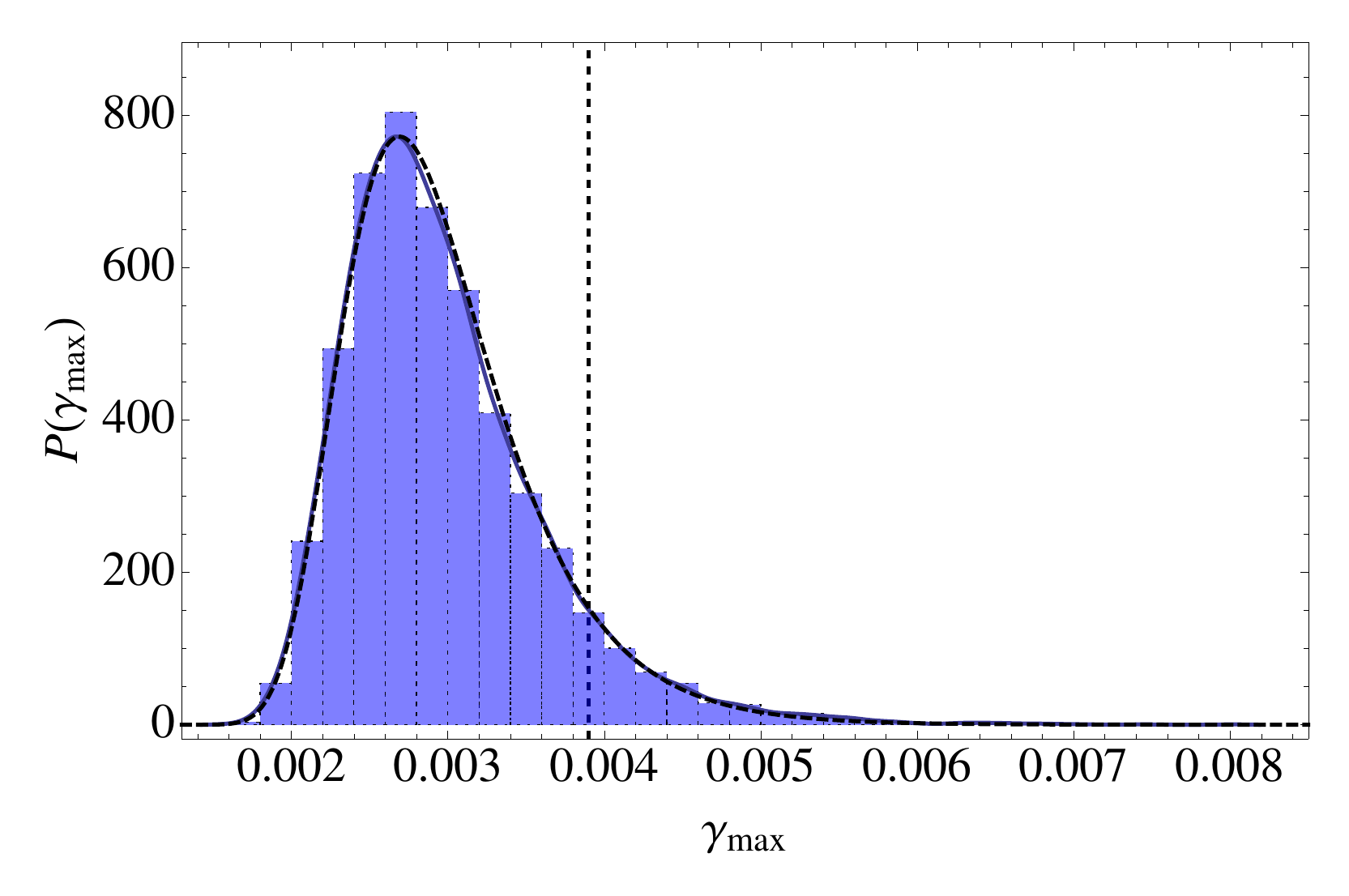}
\caption{The probability density distribution of maximum cluster memberships found in $10,000$ homogeneous random simulations, shown in terms of the variable $\gamma_\mathrm{max}=k_\mathrm{max}/N$. The blue towers are a density histogram of the measured values, in bins of width $2\times10^{-4}$. The blue solid line shows a smoothed fit to the data, using a Gaussian kernel of width $10^{-4}$. The black dashed line is the best-fit form of eq.~(\ref{eq:Gumbel}). The vertical dashed line indicates the value of $\gamma_\mathrm{max}$ for the Huge-LQG.} 
\label{figure:Richness}
\end{center}
\end{figure}

We can provide a description of the probability of finding a cluster of given size in a Poisson distribution in terms of an extreme value distribution of Type I, also known as a Gumbel distribution. This has a probability density function
\beq
\label{eq:Gumbel}
P(x)\mathrm{d}x=\frac{1}{\sigma}\mathrm{e}^{-z}\mathrm{e}^{-\mathrm{e}^{-z}}\mathrm{d}x\,,
\eeq
where $z=(x-\mu)/\sigma$. This is commonly used to model the distribution of the maximum of a sample of random numbers drawn from various other distributions. Instead of using the membership value directly, we could rescale $k_\mathrm{max}$ by the total number of points in each set, $\gamma_\mathrm{max}=k_\mathrm{max}/N$. The distribution of this quantity is less dependent than that of $k_\mathrm{max}$ on the specific details of this particular simulation, such as the volume and the total number of points, and I suggest that it may be of more universal relevance, though further tests are required to confirm this. Figure~\ref{figure:Richness} shows the deduced probability density function for $\gamma_\mathrm{max}$ from the simulations, together with a plot of eq.~(\ref{eq:Gumbel}) with the best-fit parameters $\mu=2.69\times10^{-3}$ and $\sigma=4.67\times10^{-4}$, which is seen to be an extremely good description. The value of $\gamma_\mathrm{max}$ for the Huge-LQG is also indicated. The corresponding best-fit values for the distribution in the case $\beta=1$ are $\mu=6.56\times10^{-4}$ and $\sigma=8.51\times10^{-5}$. Similarly good fits are found using distributions of the form of eq.~(\ref{eq:Gumbel}) for all measured quantities from simulations.

\begin{figure}
\begin{center}
\includegraphics[width=84mm]{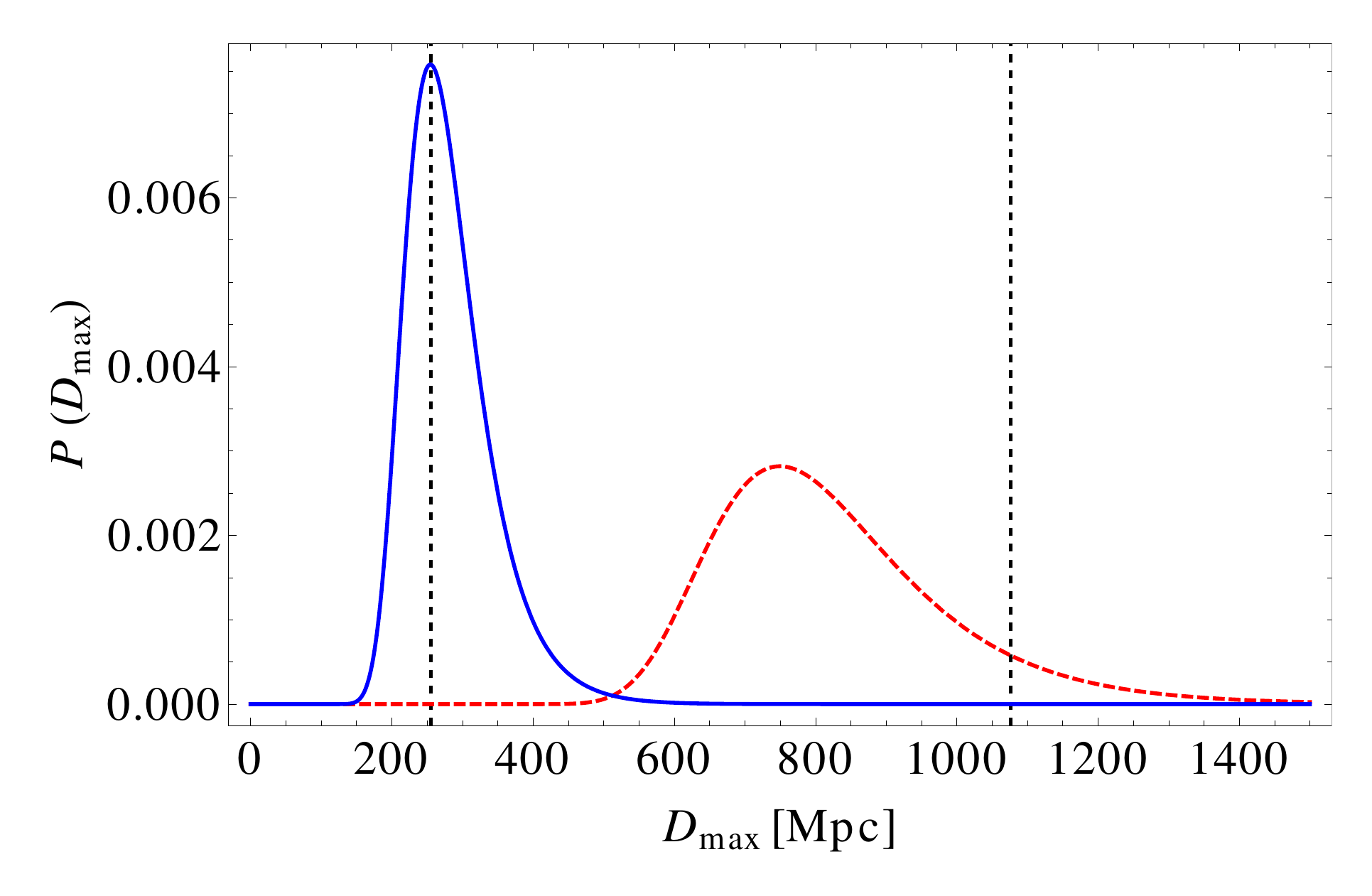}
\caption{Probability density distributions for the maximum point-to-point separation $D_\mathrm{max}$ in the largest cluster found in random simulations. The blue solid line is for linkage length $L=\bar{r}_\mathrm{nn}$, i.e. $\beta=1$. The red dashed line is for $\beta=1.33$. The vertical dashed lines show the corresponding values for the clusters found in the SCR quasar subsample.} 
\label{figure:Dmax}
\end{center}
\end{figure}

The Huge-LQG is claimed to be unusual not only because of the number of its quasar members, but also because of its spatial extent. I have chosen to quantify this in terms of the maximum separation between any two members of the cluster, $D_\mathrm{max}$. The best-fit probability density functions for $D_\mathrm{max}$ of the form of eq.~(\ref{eq:Gumbel}) are shown in Figure~\ref{figure:Dmax}, both for the case $\beta=1.33$ relevant to the Huge-LQG, and for $\beta=1$.
This clearly shows both that the hierarchical clustering algorithm can find clusters extending over hundreds of Mpc or even Gpc even in homogeneous distributions of points if the linkage length $L$ is chosen too loosely, and also that the largest clusters of quasars actually found in the DR7QSO catalogue are not significantly different to those found in homogenous random catalogues.

\begin{figure}
\begin{center}
\includegraphics[width=83mm]{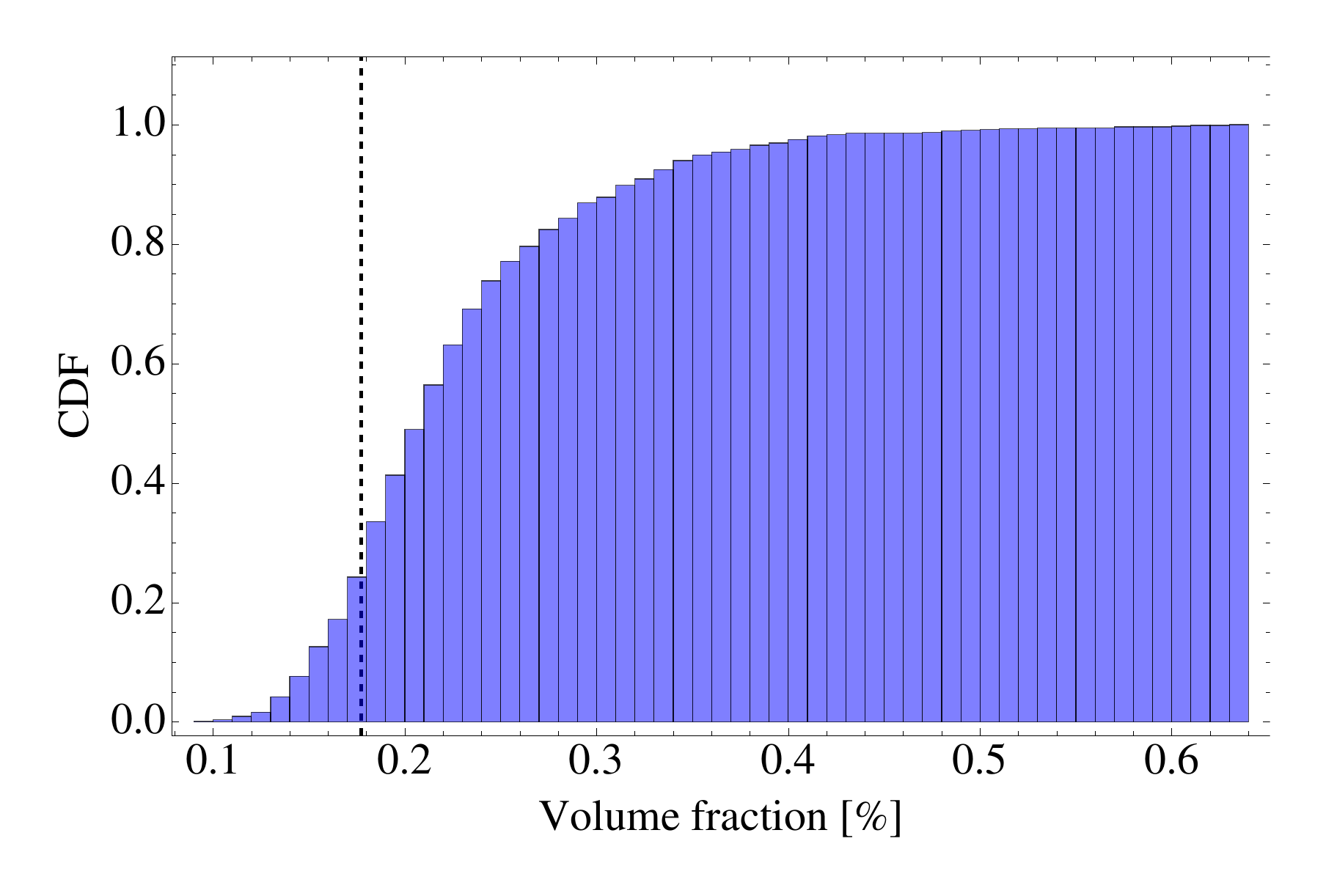}
\caption{The cumulative density function for the comoving hull volume of the set of points constituting the largest cluster found in simulations, expressed as a percentage of the total SCR volume, for only those simulated clusters which have equal or larger number of points than the Huge-LQG. The vertical line indicates the value for the Huge-LQG; roughly $25\%$ of larger clusters are also more tightly linked.} 
\label{figure:Volfrac}
\end{center}
\end{figure}

Finally, Figure~\ref{figure:Volfrac} shows the cumulative distribution function for the convex hull volume of the largest cluster found in the simulation, expressed as a percentage of the total volume of the SCR, for only those cases (amounting to $\sim8.5\%$ of the total) where the largest cluster contained as many points as the Huge-LQG or more. It can be seen that roughly $25\%$ of such clusters, despite having more members, occupy a smaller volume and so are more tightly linked than the Huge-LQG.  Since it compares like with like, this is a more appropriate statistical measure of the unlikeliness of finding such a structure in random noise than that performed by \citet{Clowes:2012pn}.

\section{Conclusion}
\label{section:conclusion}

The question of whether the observed distribution of objects in the Universe is consistent with the assumption of large-scale homogeneity and isotropy is a very important one because of the central role the cosmological principle plays in almost all theoretical models. If evidence for the violation of homogeneity were to be found, this would constitute a serious problem for the standard cosmological model. The claim that quasar structures in the DR7QSO catalogue challenged the cosmological principle \citep{Clowes:2011eb,Clowes:2012pn} therefore needed to be taken seriously, and an investigation of this issue was the major objective of this paper.  

However, this claim has been shown to be mistaken, on several counts. Firstly, as was argued in Section~\ref{section:homogeneity}, the existence of individual structures in a catalogue, even if they are of Gpc sizes, cannot be used to make inferences about the homogeneity or otherwise of the catalogue as a whole. The homogeneity of a catalogue is established by different methods to those used to identify structures, and direct comparisons between the length scales involved are not possible. This is of course not a completely new insight, but clarification of the point was evidently required.

In fact, because of its very large volume, the DR7QSO catalogue can be used to test the homogeneity of the quasar distribution out to much larger scales than probed by previous studies with other surveys, and making fewer \emph{a priori} assumptions of the homogeneity that is to be tested. I used the standard fractal analysis technique to show that the quasar distribution is indeed perfectly compatible with homogeneity at scales above at most $R_\mathrm{H}\sim130\,h^{-1}$~Mpc.

The evident homogeneity of the quasar distribution at scales far smaller than the sizes of the clusters claimed to have been detected also raises questions about the algorithm used for this detection. The operation of this algorithm depends crucially on the value of the maximum linkage length $L$. The detection of the Huge-LQG and other claimed quasar structures relied on a value $L=100$~Mpc, which is significantly larger than the mean nearest-neighbour separation for the quasars. The justification for this provided by \citet{Clowes:2011eb} is that smaller values increase the probability of failing to detect existing structures; however, the opposite is also true -- increasing $L$ increases the probability of false positive detections. This probability can be quantified by the use of simulations of homogeneous Poisson distributions of points, occupying the same volume as the quasar sample and with the same mean density. Analysis of the operation of the clustering algorithm on 10,000 such simulations shows that clusters that are larger than the claimed quasar structures -- both in number of members and spatial extent -- are quite common.

In general when using an algorithmic approach to identify clusters of points in a distribution, one must employ some criterion in order to decide whether the results obtained correspond to `real' structures in the Universe, or are merely artifacts of the algorithm. One possible criterion is theoretical: if there is a good reason to believe that the points in the cluster are in fact gravitationally bound, for instance, or if its properties match those of structures that are expected to exist in the real Universe, it may be regarded as real. Alternatively, to assess unusual clusters which do not conform to theoretical expectation, the relevant criterion is whether they are unlikely to have arisen purely from noise.

Since the linkage length used to identify the Huge-LQG is so large, there is no reason I know of to believe that it forms a gravitationally bound structure. Certainly no real structures of such size are expected in the standard cosmology. On the other hand, when using this linkage length the clustering algorithm often finds such extended structures even in pure Poisson noise. It therefore appears that the Huge-LQG fails to satisfy either criterion, and so its interpretation as a `structure' is highly questionable. This conclusion is even more applicable to the other slightly smaller quasar groups whose existence has also been claimed \citep[e.g.][]{Clowes:1991,Clowes:2011eb}.

Finally, it is worth noting that a similar situation arose recently with respect to the Sloan Great Wall. Based on the use of a very similar clustering algorithm for its identification, \citet{Sheth:2011gi} argued that the SGW was very unlikely in a $\Lambda$CDM cosmology, but \citet{Park:2012dn} found that the algorithm often identified even bigger structures in simulations. We should regard this as a reminder not to trust inferences based on rare structures found using such algorithms in the absence of a proper quantification of their action on simulated distributions. At the very least, one needs to use Poisson distributions to test the null hypothesis, as done in this paper. However, if the linkage length $L$ used is of order the scale of clustering in $\Lambda$CDM ($\sim10$~Mpc), this will not be enough and full $N$-body simulations in $\Lambda$CDM are required.

\section{Acknowledgements}
I thank Shaun Hotchkiss for discussions that led to this project and for involvement in it at an early stage. Additional thanks are due to Max Atkin and Dominik Schwarz for helpful discussions. I acknowledge support from the Sofja Kovalevskaja program of the Alexander von Humboldt Foundation. 

This research has used the SDSS DR7QSO catalogue of \citet{Schneider:2010hm}. Funding for the SDSS and SDSS-II has been provided by the Alfred P. Sloan Foundation, the Participating Institutions, the National Science Foundation, the U.S. Department of Energy, the National Aeronautics and Space Administration, the Japanese Monbukagakusho, the Max Planck Society, and the Higher Education Funding Council for England. The SDSS Web Site is http://www.sdss.org/.

The SDSS is managed by the Astrophysical Research Consortium for the Participating Institutions. The Participating Institutions are the American Museum of Natural History, Astrophysical Institute Potsdam, University of Basel, University of Cambridge, Case Western Reserve University, University of Chicago, Drexel University, Fermilab, the Institute for Advanced Study, the Japan Participation Group, Johns Hopkins University, the Joint Institute for Nuclear Astrophysics, the Kavli Institute for Particle Astrophysics and Cosmology, the Korean Scientist Group, the Chinese Academy of Sciences (LAMOST), Los Alamos National Laboratory, the Max-Planck-Institute for Astronomy (MPIA), the Max-Planck-Institute for Astrophysics (MPA), New Mexico State University, Ohio State University, University of Pittsburgh, University of Portsmouth, Princeton University, the United States Naval Observatory, and the University of Washington.

\bibliography{../../refs.bib}
\bibliographystyle{mn2e}

\label{lastpage}
\end{document}